# Resistive Plate Chambers for brain PET imaging and particle tracking and timing (TOF-tracker)

Paulo Fonte[1,2,3,*], Luís Lopes[2], Filomena M. C. Clemêncio[4], Miguel Couceiro[1,2], Susete Fetal[1,2], Custódio F. M. Loureiro[5], Jan Michel[6], João Saraiva[2], Michael Traxler[7], Antero Abrunhosa[3,8], Alberto Blanco[2], Miguel Castelo-Branco[3,8]

[1] Coimbra Institute of Engineering, Polytechnic University of Coimbra, Rua Pedro Nunes - Quinta da Nora, 3030-199 Coimbra, Portugal.

[2] LIP - Laboratory of Instrumentation and Experimental Particle Physics, Coimbra, Portugal

[3] Coimbra Institute for Biomedical Imaging and Translational Research (CIBIT), University of Coimbra - Coimbra, Portugal

[4] Escola Superior de Tecnologia da Saúde do Porto—IPP, Vila Nova de Gaia, Portugal

[5] Department of Physics, University of Coimbra, 3004-516 Coimbra, Portugal

[6] Goethe Univ, Inst. Kernphys, D-60438 Frankfurt, Germany

[7] GSI Helmholtzzentrum für Schwerionenforschung GmbH, D-64291 Darmstadt, Germany

[8] ICNAS Pharma, Institute for Nuclear Sciences Applied to Health, University of Coimbra - Coimbra, Portugal

* Corresponding author: Instituto Superior de Engenharia de Coimbra, Rua Pedro Nunes, 3030-199, Coimbra, Portugal.

**Abstract**

In this work we explore readout architectures for the simultaneous high-resolution timing and bidimensional tracking of charged particles with Resistive Plate Chambers (TOF-tracker) and for the accurate detection of gamma rays for PET imaging.

On 625 cm$^2$ of active area we obtained a time resolution of 61 ps σ and bidimensional position resolution below 150 μm σ for the tracking and timing of charged particles from cosmic rays. An intrinsic precision of 0.49 mm FWHM was determined for the localization of a small β$^+$ source via the detection of its annihilation radiation.

# Introduction

Resistive Plate Chambers (RPC) are widely used gaseous detectors, most commonly for the detection of charged particles in High Energy Physics experiments, including large time-of-flight detectors and trigger detectors with coarse positioning capability. Applications have been found also in cosmic ray research.

There is in principle no reason why the position accuracy of RPCs should be inferior to other types of particle detector and we have proposed and demonstrated long ago the "TOF-tracker" concept, combining sub-100 ps timing and sub 0.1 mm position resolutions, however in a small area [1]. Other instances of the concept were studied, some featuring much larger areas, but generally with inferior time or position resolutions [2]-[5].

In here we extend the demonstration of a high position and time resolution TOF-tracker device to an area of 900 cm$^2$, of which 625 cm$^2$ were considered as active, and study its application to charged particle tracking and Positron Emission Tomography (PET) imaging.

Three experiments were performed. Experiments 1 and 2 concerned a 4-layer telescope housed in two different gas enclosures ("heads") whose distance could be varied. Experiment 1 was about the precision of localization of a small $\beta^+$ source via the detection of its annihilation radiation, which depends on the distance between the detecting planes. Experiment 2 was carried on the same setup with the 4 planes approximately equidistant and determined the time and position precision for charged particles (cosmic rays).

These two experiments were performed with a simplified time readout method featuring only two timing channels per layer [6], while the standard time readout method requires one timing channel at each end of each readout strip. To compare both methods, Experiment 3, concerned only one plane pair but with both planes readout in the standard way.

# Detectors and setup

Each independent plane (see Figure 1) of the telescopes was constituted by two 5-gap multigap RPCs of the type described in [6] and readout in a similar way.

In between the two RPCs of the plane was placed the X-coordinate readout electrode. The time readout is combined with the X-position readout in the way described in [1], forming ten electrical structures that are electrically equivalent to readout strips.

Both exterior faces of the plane were covered by the Y-coordinate readout electrodes. The corresponding signals from the top and bottom electrodes were connected together to the same amplifier channel, 48 channels per coordinate.

The gas enclosures, each containing two planes and the corresponding readout electronics, were also of the type described in [6]. The gas was a mixture of R134a (98%) and SF6 (2%) at a total flow of 20 cc/minute.

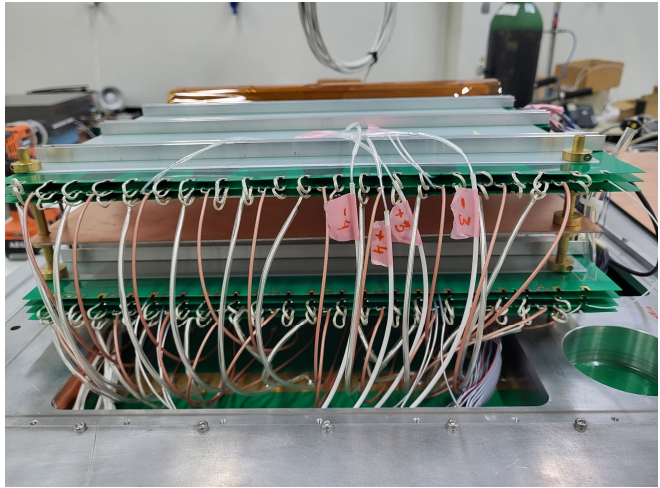

Figure 1 – View of the detector stack in one of the heads. The brown cables are the timing connections to the central X electrodes, which are surrounded by 2 RPCs and external Y electrodes. The other cables are ground or high-voltage (those labelled) connections. The flat-cable connections to the 48 readout strips/electrode are on the right-hand side and cannot be seen.

## Methods

### *Experiment 1: localization of a β⁺ source*

To evaluate the intrinsic precision of localization of gamma-ray interactions in the RPC a PET-like experiment was performed. A small $^{22}$Na source with 0.2 mm diameter enclosed in PMMA was placed between the 2 heads, each head housing two RPC layers. A schematic representation of the experimental setup can be seen in Figure 2.

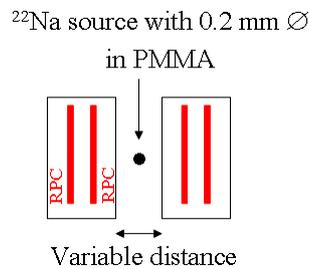

Figure 2 – Schematic representation of the experimental setup for Experiment 1.

A large set of lines-of-response (LORs), connecting the detected points on each plane, was collected and the precision of the localization of the source was estimated as the radial width of the distribution of the intersection of the LORs with the plane of least confusion (smaller width) parallel to the detecting planes. The location of the plane of least confusion was experimentally determined for each layer pair and each distance between heads.

A complication of this measurement is that the angle between the two photons arising from the annihilation of the positron is not exactly 180º, with a distribution about this value that depends on the material on which the positron annihilates. For the material housing the source this was measured to be 8.4 mrad FWHM [7]. For a point source at the center of the field-of-view and using the procedure described in the previous paragraph this corresponds to a FWHM deviation of 2.1 mm per meter of distance between the detecting planes.

Trying to disentangle this effect from the intrinsic resolution we performed the measurements as a function of the distance between detecting planes.

## *Experiment 2: 4-plane charged-particle telescope*

For this experiment we used a simplified channel-saving time readout scheme, illustrated in Figure 3. Each of the 10 time readout strips is connected on a single end to an amplifier-comparator circuit (see [6] for details). The LVDS outputs of the comparators are wired-ORed into a differential 100 Ω line that is sensed on both ends by comparators followed by TDCs (see [6]). The rationale for such readout was that any position dependencies of the time could be compensated by software, as precise bidimensional coordinates are available.

The precision of the time measuring system was estimated by pulsing synchronously the front-end electronics and measuring the time difference between a pair of TDCs in different gas enclosures, yielding a value of 93 ps σ per TDC channel.

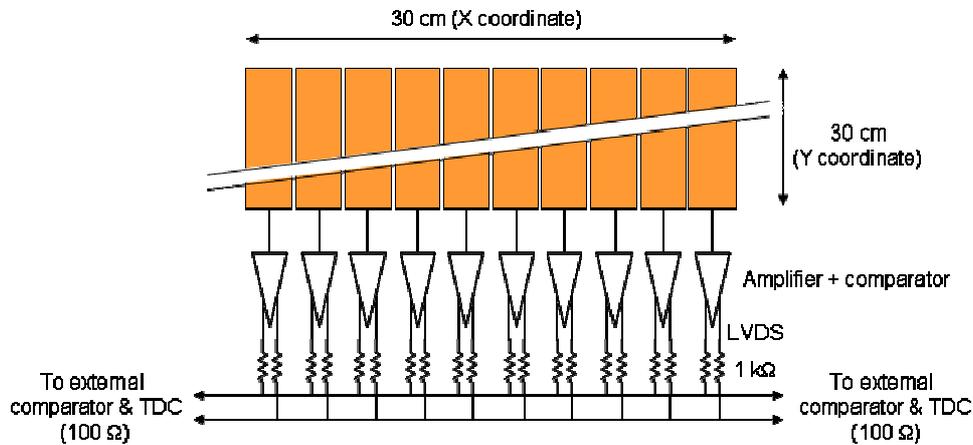

Figure 3 – Schematic of the simplified time readout used in experiment 2. The strips depicted are not physical, but are aggregated from the position readout strips as described in [1].

Data was recorded when any of the two planes on both enclosures had a valid time signal. The applied voltage was 15.6 kV per RPC.

The position of the avalanches in each coordinate (X and Y) of each plane was determined by charge interpolation between the 48 position readout strips.

To avoid edge effects, the fiducial area of the telescope was defined as the central 25 cm in each coordinate. To avoid excessive parallax errors due to the thickness of the planes only tracks with inclination relative to the vertical smaller than 2 degrees were analysed.

The position resolution in each coordinate was estimated by fitting a straight line to the four measured points in the minimum least squares sense. This procedure generates residuals whose dispersion is not equal to the precision of the coordinates in each plane. To uniformise the results the residuals were weighted by fixed factors so that the final numbers ("weighted fit residuals") are equal to the expected precision of each layer for the case when all layers have the same precision. These weights were determined by applying the same fitting procedure to synthetic data of known standard deviation per layer.

To determine the relevant standard deviations the distributions were fitted by a Gaussian function but restricting the fit to ±1.5 σ to avoid the contribution of any non-gaussian tails. This procedure was used for all such measurements.

It is clear that with this setup, all the layers having approximately the same precision, it is not possible to determine individual precisions by layer, but only a global value for the set of layers.

Supplementing the mechanical alignment, the position in space of each layer was algorithmically determined by minimization of the fit residuals.

Concerning the time resolution, the final standard deviation figures were determined by fitting the distribution of the time differences between each pair of layers following the same procedure as described above. Then it was assumed that both layers offered the same precision, implying that the precision by layer was the determined standard deviations divided by $\sqrt{2}$.

The time difference values were corrected empirically for correlations with the avalanche charge and position.

## *Experiment 3: 2-plane telescope*

In this experiment the time readout was upgraded from the simplified scheme discussed above to the more standard readout schematized in Figure 3. Owing to some limitations in the available hardware this could only be implemented in a single gas enclosure (two sensitive planes), so no simultaneous position resolution was available.

The final standard deviation figures were determined as described in the previous section.

The time difference values were corrected empirically for correlations with the avalanche charge and position, but there was negligible correlation with position.

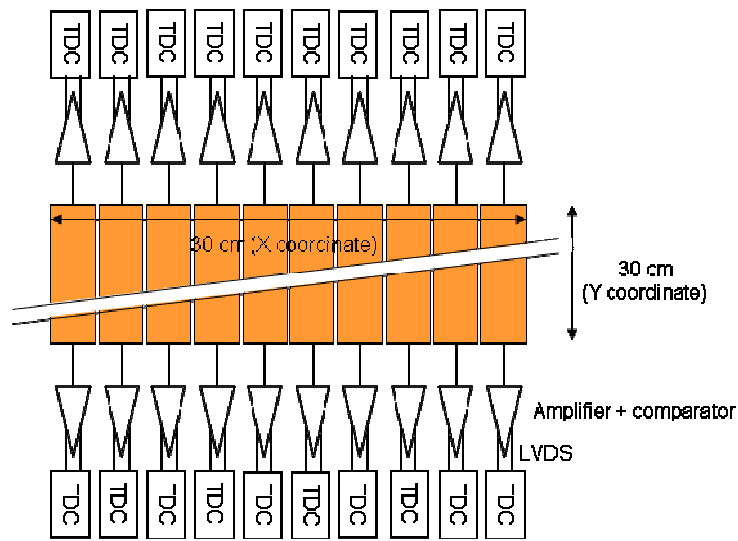

Figure 4 – Schematic of the time readout used for experiment 3.

It was determined by switching off the high voltage from one of the layers that the crosstalk between layers was on the level of a few percent, therefore statistically insignificant.

# Results

## *Experiment 1: localization of a β⁺ source*

The experimental results are presented in Figure 5 (circles), along with some quantities useful for their interpretation.

It can be appreciated that the width of the distribution grows almost linearly with the distance between detecting planes, with a value extrapolated to zero distance of 0.494 mm, being this value assumed to be the intrinsic localization precision. This would correspond to a precision per layer of $0.494 \times 2 / \sqrt{2} = 0.699$ mm.

It is also apparent that this dependency with the distance cannot be explained solely by the non-colinearity of the photons (dashed lines) and it is also not very dependent on the angular inclination of the photons (polar angle cut) which could cause some parallax effect. Therefore some further work remains to be done concerning the origin of the effect.

For the widest distance, about 350 mm, essentially corresponding to the device described in [6], the width was 1.227 mm while the corresponding image had resolution below 1 mm. Using this scaling one might estimate that the image resolution corresponding to zero distance would be 0.494/1.22×1=0,404 mm.

Of course, in this discussion we are neglecting the 0.2 mm diameter of the radioactive source.

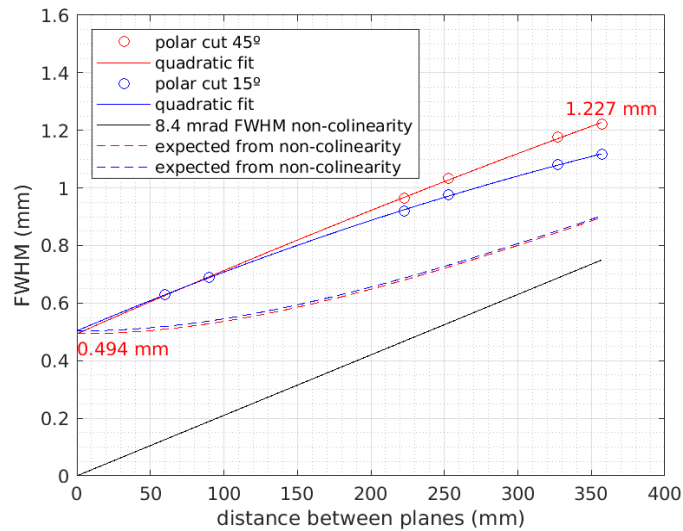

Figure 5 - The measured FWHM widths of the distribution of the intersection of the LORs with the plane of least confusion as a function of the distance between these planes. The presumable effects of non-colinearity are superimposed.

## *Experiment 2: 4-plane charged-particle telescope*

The distribution of the weighted fit residuals determined from experiment 2 are shown in Figure 6, along with the corresponding Gaussian fits. The average standard deviation values are 130 µm and 128 µm respectively for the X and Y coordinates.

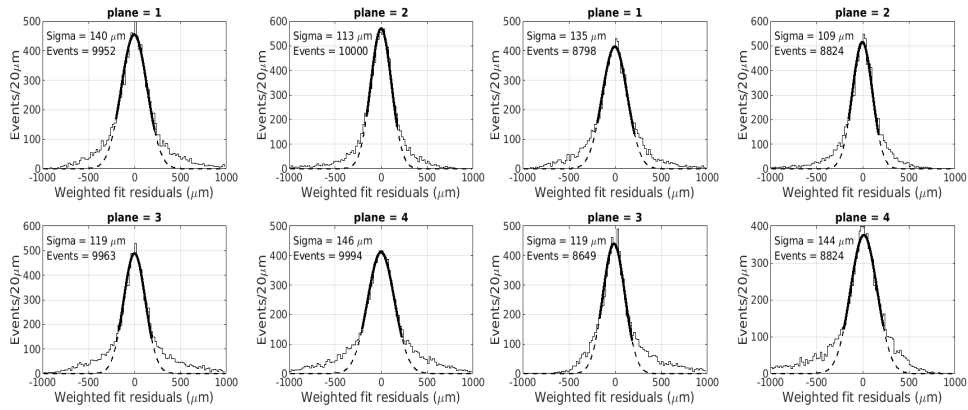

Figure 6 – Gaussian fits to the distribution of the weighted fit residuals for the X (left panel) and Y (right panel) coordinates and each of the four telescope planes.

Additionally, we made the exercise of injecting by software a noise in each channel equivalent to the charge pedestal measured by pulsing the entrance of the amplifiers during the data taking (therefore in fully realistic conditions). When the injected noise was 10-fold the measured noise value the resolution figures only worsened by 32%, suggesting that the measured resolution is not dominated by electronic noise but by other factors.

The time precision between pairs of layers determined according to the procedure described above (an illustration is shown in Figure 7) is detailed in Table 1, both for the case when the average of both TDC times were used (see Figure 3) and when only one TDC was used. In both cases the time difference values were corrected empirically for correlations with the avalanche charge and position, but for the former case there was negligible correlation with position.

Table 1- Measured timing precision (ps) between pairs of layers, divided by $\sqrt{2}$. The upper triangle corresponds to the values obtained by using both TDC values available in each layer (see Figure 3) and the lower one to using only one of the TDC values. The average values are, respectively, 222 ps and 273 ps.

|        | layers |     |     |     |
|--------|--------|-----|-----|-----|
| layers | 1      | 2   | 3   | 4   |
| 1      |        | 160 | 253 | 213 |
| 2      | 207    |     | 240 | 198 |
| 3      | 297    | 278 |     | 268 |
| 4      | 277    | 275 | 302 |     |

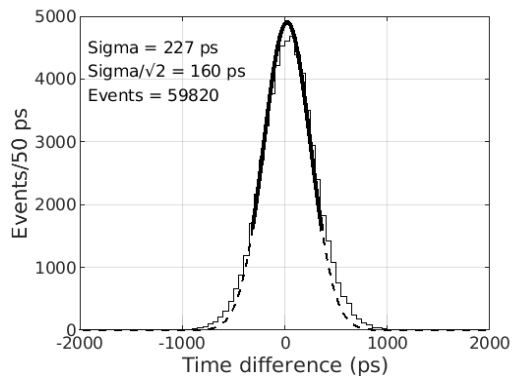

Figure 7 – Example of the time difference distribution between layers 1 and 2 and respective fit.

## *Experiment 3: 2-plane telescope*

The timing precision measured between the two layers in accordance with the procedure described above is shown in Figure 8 as a function of the applied voltage per gap, ranging from 82 to 61 ps.

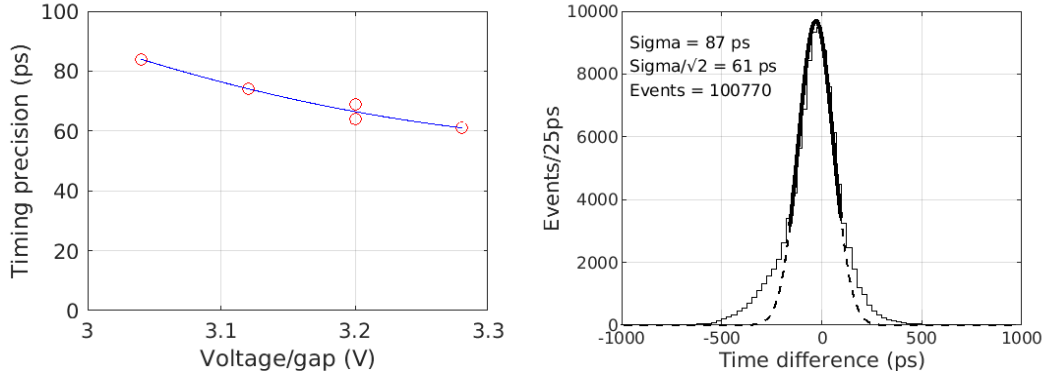

Figure 8 – Left panel: timing precision as a function of the voltage applied by gap. Right panel: illustration of the fitting procedure for the point at higher voltage.

## Conclusions

A 4-plane telescope with active area of 25×25 cm$^2$ was equipped with position- and time-sensitive electronics and tested with cosmic rays. It was observed an average position resolution per plane of 130 μm σ in the X direction and of 128 μm σ in the Y direction. These values are not dominated by electronic noise, but, presumably, by systematic effects. For this experiment it was used a simplified time readout, yielding an average time resolution of 222 ps σ.

Using the same device, the localization precision of a small β$^+$ source from its annihilation radiation was investigated as a function of the distance between detecting planes, yielding an intrinsic precision of 0.49 mm FWHM when extrapolated to the distance zero.

In a third experiment with a 2-plane telescope the position readout was kept unchanged and the time readout was improved by individually reading both ends of the 10 timing readout strips formed over the active area of each plane. In this case the timing resolution per plane was 61 ps σ.

## Acknowledgements

This work was financed by the European Union and ANI - Agência Nacional de Inovação, S.A. via the programs PT2020 and COMPETE2020 (project POCI-01-0247-FEDER-039808) and by Fundação para a Ciência e Tecnologia (projects CERN-FIS-INS-0009-2019 and CERN/FIS-INS/0006/2021).

## References

[1] Blanco, P. Fonte, L. Lopes, P. Martins, J. Michel, M. Palka, M. Kajetanowicz, G. Korcyl, M. Traxler, R. Marques, TOFtracker: gaseous detector with bidimensional

tracking and time-of-flight capabilities, J. Inst. 7 (2012) P11012–P11012. https://doi.org/10.1088/1748-0221/7/11/P11012.

[2] G. Aielli, R. Cardarelli, L.D. Stante, B. Liberti, L. Paolozzi, E. Pastori, R. Santonico, The RPC space resolution with the charge centroid method, J. Inst. 9 (2014) C09030–C09030. https://doi.org/10.1088/1748-0221/9/09/C09030.

[3] P. Assis, A. Bernardino, A. Blanco, F. Clemêncio, N. Carolino, O. Cunha, M. Ferreira, P. Fonte, L. Lopes, C. Loureiro, R. Luz, L. Mendes, J. Michel, A. Neiser, A. Pereira, M. Pimenta, R. Shellard, M. Traxler, A large area TOF-tracker device based on multi-gap Resistive Plate Chambers, J. Inst. 11 (2016) C10002–C10002. https://doi.org/10.1088/1748-0221/11/10/C10002.

[4] X.L. Chen, Y. Wang, G. Chen, D. Han, B. Guo, F. Wang, Y. Yu, B. Wang, Q. Zhang, Y. Li, MRPC technology for muon tomography, J. Inst. 15 (2020) C12001–C12001. https://doi.org/10.1088/1748-0221/15/12/C12001.

[5] R. Uda, F. Hayashi, N. Tomida, W.-C. Chang, M.-L. Chu, C.-Y. Hsieh, P.-J. Lin, K. Miyamoto, H. Noumi, A. Sakaguchi, K. Shirotori, M. Tokuda, T. Toda, Y. Yamamoto, Development of a precise time and position resolution TOF-tracker MRPC for the π20 beam line at J-PARC, Nucl. Instrum. and Meth. in Phys. Res. A 1056 (2023) 168580. https://doi.org/10.1016/j.nima.2023.168580.

[6] P. Fonte, L. Lopes, R. Alves, N. Carolino, P. Crespo, M. Couceiro, O. Cunha, N. Dias, N.C. Ferreira, S. Fetal, A.L. Lopes, J. Michel, J. Moreira, A. Pereira, J. Saraiva, C. Silva, M. Silva, M. Traxler, A. Abrunhosa, A. Blanco, M. Castelo-Branco, M. Pimenta, An RPC-PET brain scanner demonstrator: First results, Nucl. Instrum. and Meth. in Phys. Res. A 1051 (2023) 168236. https://doi.org/10.1016/j.nima.2023.168236.

[7] Y.C. Jean, H. Nakanishi, L.Y. Hao, T.C. Sandreczki, Anisotropy of free-volume-hole dimensions in polymers probed by positron-annihilation spectroscopy, Phys. Rev. B 42 (1990) 9705–9708. https://doi.org/10.1103/PhysRevB.42.9705.